\documentclass[prd,twocolumn,showpacs,amsmath,amssymb]{revtex4}
\usepackage{graphicx}

\begin{document}

\title{\boldmath
Measurement of Cross Sections for $D^0 {\bar D}^0$
and $D^+D^-$ Production in $e^+e^-$ Annihilation 
at $\sqrt{s}=3.773$ GeV}
\author{
M.~Ablikim$^{1}$, J.~Z.~Bai$^{1}$, Y.~Ban$^{10}$, 
J.~G.~Bian$^{1}$, X.~Cai$^{1}$, J.~F.~Chang$^{1}$, 
H.~F.~Chen$^{15}$, H.~S.~Chen$^{1}$, H.~X.~Chen$^{1}$, 
J.~C.~Chen$^{1}$
% ...
\\(BES Collaboration)\\ 
}
\vspace{0.2cm}
\affiliation{
\begin{minipage}{145mm}
$^{1}$ Institute of High Energy Physics, Beijing 100039, People's Republic of China\\
$^{2}$ China Center for Advanced Science and Technology, 
Beijing 100080, People's Republic of China\\
$^{3}$ Guangxi Normal University, Guilin 541004, People's Republic of China\\
$^{4}$ Henan Normal University, Xinxiang 453002, People's Republic of China\\
$^{5}$ Huazhong Normal University, Wuhan 430079, People's Republic of China\\
$^{6}$ Hunan University, Changsha 410082, People's Republic of China\\
$^{7}$ Liaoning University, Shenyang 110036, People's Republic of China\\
$^{8}$ Nanjing Normal University, Nanjing 210097, People's Republic of China\\
$^{9}$ Nankai University, Tianjin 300071, People's Republic of China\\
$^{10}$ Peking University, Beijing 100871, People's Republic of China\\
$^{11}$ Shandong University, Jinan 250100, People's Republic of China\\
$^{12}$ Shanghai Jiaotong University, Shanghai 200030, People's Republic of China\\
$^{13}$ Sichuan University, Chengdu 610064, People's Republic of China\\
$^{14}$ Tsinghua University, Beijing 100084, People's Republic of China\\
$^{15}$ University of Science and Technology of China, Hefei 230026, 
    People's Republic of China\\
$^{16}$ Wuhan University, Wuhan 430072, People's Republic of China\\
$^{17}$ Zhejiang University, Hangzhou 310028, People's Republic of China\\
\vspace{0.4cm}
\end{minipage}
}
\vspace{0.2cm}

\begin{abstract}
The cross sections for $D^0 {\bar D}^0$ and $D^+D^-$
production at 3.773 GeV have been measured with BES-II 
detector at BEPC. These measurements are
made by analyzing a data sample of about $17.3$ $\rm pb^{-1}$ 
collected at the center-of-mass energy of 3.773 GeV. Observed cross sections
for the charm pair production are radiatively corrected to obtain the
tree level cross section for $D\bar D$ production. 
A measurement of the total
tree level hadronic cross section is obtained from the tree level
$D \bar D$ cross section and an extrapolation
of the $R_{uds}$ below the open charm threshold.  
\end{abstract}

\maketitle


\section{Introduction}

Around the center-of-mass energy of 3.770 GeV,
the $\psi(3770)$ resonance is produced in $e^+e^-$ annihilation,
and open charm pairs, $D^0 {\bar D}^0$ and 
$D^+D^-$, are mainly produced in $\psi(3770)$ decays. 
So the measurement of the cross sections for $D^0 {\bar D}^0$ and
$D^+D^-$ production at an energy point around 3.770 GeV is
very important in the understanding of $\psi(3770)$ decays.
A coupled-channel model~\cite{Eichten} predicts that the cross section for
$\psi(3770)$ production is about $3~ \rm nb$ and 
that the $\psi(3770)$ decay exclusively into $D^0 {\bar D}^0$ and $D^+D^-$.
Experimental results on the measurement of the 
$D^0 {\bar D}^0$, $D^+D^-$ and
$D\bar D$ cross sections can be used to test
the theoretical prediction.
Also the measured values of the cross sections can be used to determine the
branching fraction for $\psi(3770) \rightarrow {\rm non} D\bar D$
using the measured cross section for $\psi(3770) \rightarrow {\rm hadrons}$
at the same energy point. The determination of the partial width
of $\psi(3770) \rightarrow {\rm non} D\bar D$ has
great interest since it would be helpful for investigating the
mixing between S and D waves in its wave function~\cite{JLRosner},
and in turn to help in developing the 
Potential Model~\cite{Eichten}. 

In addition, by adding the tree level open charm cross section 
to an extrapolation of the tree level hadronic cross section 
for the light hadron production in the region
below the open charm threshold, 
the total tree level hadronic cross section 
can be obtained~\cite{bes4030}\cite{coles}.
The tree level cross section for inclusive hadronic event production
in the $e^+e^-$ annihilation at all energies is needed 
to calculate the effects of vacuum polarization 
on the parameters of the Standard Model.
The largest uncertainty in this calculation arises from the uncertainties
in the measured inclusive hadronic cross sections in the open charm
threshold region. 
Traditionally, the tree level hadronic cross sections are measured by
counting inclusive hadronic events. 
Using the measured cross sections for the $D \bar D$ production in
the charm threshold region, we can also
measure the tree level inclusive hadronic cross sections.

\section{The BES-II Detector}
BES-II is a conventional cylindrical magnetic detector that is
described in detail in Ref.~\cite{BES-II}.  A 12-layer vertex chamber
(VC) surrounding the beryllium beam pipe provides input to the event
trigger, as well as coordinate information.  A forty-layer main drift
chamber (MDC) located just outside the VC yields precise measurements
of charged particle trajectories with a solid angle coverage of $85\%$
of $4\pi$; it also provides ionization energy loss ($dE/dx$)
measurements which are used for particle identification.  Momentum
resolution of $1.7\%\sqrt{1+p^2}$ ($p$ in GeV/$c$) and $dE/dx$
resolution of $8.5\%$ for Bhabha scattering electrons are obtained for
the data taken at $\sqrt{s}=3.773$ GeV. An array of 48 scintillation
counters surrounding the MDC measures the time of flight (TOF) of
charged particles with a resolution of about 180 ps for electrons.
Outside the TOF, a 12 radiation length, lead-gas barrel shower counter
(BSC), operating in limited streamer mode, measures the energies of
electrons and photons over $80\%$ of the total solid angle with an
energy resolution of $\sigma_E/E=0.22/\sqrt{E}$ ($E$ in GeV) and spatial
resolutions of
$\sigma_{\phi}=7.9$ mrad and $\sigma_Z=2.3$ cm for
electrons. A solenoidal magnet outside the BSC provides a 0.4 T
magnetic field in the central tracking region of the detector. Three
double-layer muon counters instrument the magnet flux return and serve
to identify muons with momentum greater than 500 MeV/c. They cover
$68\%$ of the total solid angle.

\section{Data sample and
method to determine $\sigma_{D\bar D}$}

The data used for this analysis were collected 
at the center-of-mass energy of 3.773 GeV 
with the Beijing Spectrometer~\cite{BES-II}
at Beijing Electron Positron Collider.
The total integrated luminosity of the data set
is 17.3 $\rm pb^{-1}$, which is obtained based on analysis
of large angle Bhabha scattering from the same data set.

The measurements of the cross sections for 
the $D^0 {\bar D}^0$ and
$D^+D^-$ production are made 
based on the analysis of singly tagged $D^0$ and $D^+$ events.
At the center-of-mass energy $\sqrt{s}=3.773$ GeV, the
$D^0$ (through this paper,
charge conjugation is implied) and $D^+$
are produced in pair via the process of
\begin{equation}
e^+e^- \rightarrow D^0 {\bar D}^0, D^+D^-.
\end{equation}
The total observed number $N_{D^0_{\rm tag}}$ ($N_{D^+_{\rm tag}}$)
of $D^0$ ($D^+$) meson and the observed cross section
$\sigma^{\rm obs}_{D^0 {\bar D}^0}$ 
($\sigma^{\rm obs}_{D^+D^-}$) are related as
\begin{equation}
 \sigma_{D^0{\bar D}^0}^{\rm obs} = \frac {N_{D^0_{\rm tag}}}
                       {2 \times L \times B \times \epsilon },
\end{equation}
\noindent
and
\begin{equation}
 \sigma_{D^+ D^-}^{\rm obs} = \frac {N_{D^+_{\rm tag}}}
                       {2 \times L \times B \times \epsilon },
\end{equation}

\noindent
where $L$ is the integrated luminosity of the data set used in the
analysis, $B$ is the branching fraction
for decay mode in question, 
and  $\epsilon$ is the efficiency determined from Monte Carlo for reconstruction
of this decay mode.

In the measurements of the cross sections, the singly tagged neutral
and charged $D$ mesons are 
observed in the invariant mass spectra
of the daughter particles from the $D$ decay.

\section{Data Analysis}
\subsection{Event selection}

The neutral and charged $D$ mesons are reconstructed in
the final states of $K^-\pi^+$, $K^-\pi^+\pi^+\pi^-$ and
$K^-\pi^+\pi^+$.
Events which contain at least
two reconstructed charged-tracks with good helix fits are selected.
In order to ensure well-measured 3-momentum vectors
and reliable charged-particle identification, 
the charged tracks used in the single tag analysis
are required to be within $|cos\theta|<$0.85, 
where $\theta$ is the angle with respect to beam direction.
All tracks must originate from the interaction region,
which require that the closest approach of the charged track
in $xy$ plane is less than 2.0 cm and
the $z$ position of the charged track is less than 20.0 cm.
Pions and kaons are identified by means of
TOF and $dE/dx$ measurements. 
The pion identification requires a consistency
with the pion hypothesis at a confidence level
greater than 0.1\%.
In order to reduce misidentification, the kaon candidate is
required to have a larger confidence level for a kaon hypothesis
than that for a pion hypothesis.

\subsection{Analysis of inclusive $D$ meson events}

Taking advantage of the $D\bar D$ pair production, 
we use a kinematic fit to candidate $D^0$ or $D^+$ decays
to improve the ratio of signal to noise and mass resolution
in the invariant mass spectrum.
The energy-constraint is imposed on the measured momenta of
the $D$ daughter particles via the kinematic fit to improve the
measured charged track three-momenta. 
Events with a kinematic fit probability
greater than $1\%$ are accepted. If more
than one combination satisfies
the fit probability greater than $1\%$,
the combination with the largest fit probability is retained.

The resulting distributions of the fitted masses of $Km\pi$
($m=1$, or $2$, or $3$) combinations, 
which are calculated using the fitted momentum vectors 
from the kinematic fit, are shown in figure 1.
The signals for $D^0$  and $D^+$ production are clearly observed
in the fitted mass spectra.
A maximum likelihood fit is performed to the mass spectrum, a Gaussian is
used to model the signal shape and a special function [6] is used to
describe the background shape. The $D^0$ and $D^+$ yields obtained from this
fit is given in table I.

\begin{figure}
\includegraphics[width=9.0cm,height=11.0cm]
{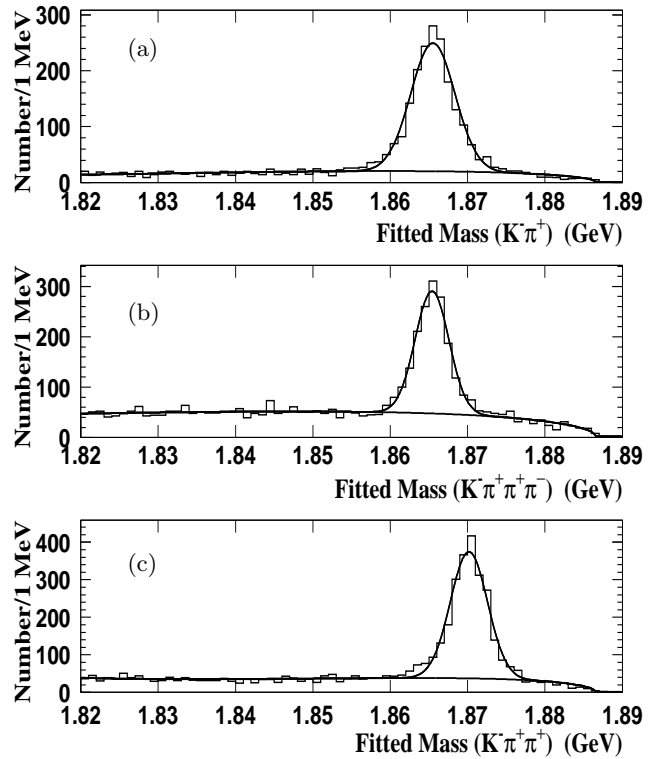}
\put(-200,280.0){(a)}
\put(-200,180.0){(b)}
\put(-200,85.0){(c)}
\caption{Distribution of the fitted masses of the $Km\pi$ (m=1,
or 2, or 3) combinations for three singly tagged modes,
where figure (a) and figure (b) are for the decay
modes of $D^0\rightarrow K^-\pi^+$ and
$D^0\rightarrow K^-\pi^+\pi^+\pi^-$, respectively,
and the figure (c) is for the decay mode of  
$D^+\rightarrow K^-\pi^+\pi^+$.
}
\end{figure}
\begin{table}
\caption{Singly tagged $D^0$ and $D^+$ samples. Where the $M_{\rm fit}$ is the fitted
mass of singly tagged $D$ meson, the $N^{\rm obs}_{D_{\rm tag}}$ is the observed number 
of singly tagged $D$ meson and the $N_{D_{\rm tag}}$ is the ''true'' number of the
singly tagged $D$ meson after correcting the contamination 
from other decay modes.
}
\begin{center}
\begin{tabular}{cccc} \hline \hline
 Tag Mode  & $M_{\rm fit}$~ [\rm MeV/$c^2$] & $N^{\rm obs}_{D_{\rm tag}}$ &
$N_{D_{\rm tag}}$ \\ 
\hline
$K^-\pi^+$ & $1865.5\pm 0.1$ & $1642.8 \pm 49.9$ & $1627.4 \pm 49.9$    \\
$K^-\pi^+\pi^+\pi^-$ & $1865.4\pm 0.1$ & $1327.2 \pm 48.6$ & $1299.1\pm 48.6$ \\
$K^-\pi^+\pi^+$ & $1870.2\pm 0.1$ & $2029.3 \pm 57.4$ & $2010.8\pm 57.4$    \\
\hline \hline
\end{tabular}
\end{center}
\end{table}
The same $Km\pi$ combinations
from other decay modes can also pass 
the above selection criteria and the distributions of the fitted masses
of the combinations are with a small peak around the masses of $D$ mesons.
The rates of the contaminations
are evaluated to be 0.0094, 0.0212 and 0.0091 for the $D^0 \rightarrow
K^-\pi^+$, $D^0 \rightarrow K^-\pi^+\pi^+\pi^-$ and 
$D^+ \rightarrow K^-\pi^+\pi^+$, respectively. 
After correcting the observed numbers of the singly tagged $D$
events for these combinations, the ''true'' numbers of the $D$
signal events for the three singly tagged $D$ modes are obtained
to be $1627\pm 50$, $1299 \pm 49$ and $2011 \pm 57$. 
Table I summarizes the results of the inclusive $D$ analysis.

\section{Observed cross sections for $D^0 {\bar D}^0$ and $D^+D^-$ production}
\subsection{Monte Carlo Efficiency}

To estimate the reconstructed efficiencies of
$D^0 \rightarrow K^-\pi^+$,
$D^0 \rightarrow K^-\pi^+\pi^+\pi^-$ and
$D^+ \rightarrow K^-\pi^+\pi^+$, the Monte Carlo samples
of $D\bar D$ production and decays are generated 
according to equation (1), where the ratio of the
neutral over the total $D\bar D$ production cross section
is set to be 0.58. Both $D$ and $\bar D$ mesons 
decay to all possible modes according to the branching fractions
quoted from PDG\cite{PDG02}. The generated
events are simulated with a GEANT based Monte Carlo package. All
decay processes which contribute to the decay modes in question are
considered in estimating the efficiencies. Detailed Monte Carlo studies
give the efficiencies to be $(35.26\pm0.19)\%$,  $(13.73 \pm 0.09)\%$ 
and  $(25.00\pm 0.13)\%$ for the reconstruction of
$D^0 \rightarrow K^-\pi^+$, $D^0 \rightarrow K^-\pi^+\pi^+\pi^-$ and 
$D^+ \rightarrow K^-\pi^+\pi^+$ decay modes, 
respectively. 

\subsection{Observed Cross Sections}

   Inserting the number of the singly tagged $D$ events and the
efficiencies for each of the three decay modes,
the $\sigma_{D\bar D}\times B$
of $D^0{\bar D}^0$ and $D^+D^-$ are obtained 
and the results are shown in table II.
The first error is statistical and second systematic which arise from
the uncertainty in the measured luminosity ($\sim 3\%$), 
tracking ($\sim 2\%$ per track), particle identification ($\sim 0.5\%$/track),
kinematic fit ($\sim 1\%$), fitting parameters ($\sim 3\%$)
and Monte Carlo statistics ($\sim 0.6\%$).
The total systematic uncertainty is obtained 
by adding all systematic uncertainties in quadrature.
\begin{table}
\caption{Summary of the observed cross section times branching fraction.}
\begin{center}
\begin{tabular}{cc}
\hline \hline
Mode  & $\sigma_{D\bar D}\times B$ [nb]  \\
\hline
$D^0 \rightarrow K^-\pi^+$           &
   $0.133\pm0.004\pm0.008$ \\
$D^0 \rightarrow K^-\pi^+\pi^+\pi^-$ &
    $0.273\pm0.010\pm0.025$ \\
$D^+ \rightarrow K^-\pi^+\pi^+$ &
     $0.233\pm0.007\pm0.018$ \\
\hline \hline
\end{tabular}
\end{center}
\end{table}
\begin{table}
\caption{A comparison of $\sigma_D \times B$ measured by
this experiment, MARK-I and MARK-II experiments.}
\begin{center}
\begin{scriptsize}
\begin{tabular}{cccc}
\hline \hline
     & $\sigma_D\times B$ [nb] & $\sigma_D\times B$ [nb]  & 
$\sigma_D\times B$ [nb] \\
Mode & (This experiment)  & (MARK-II) &(MARK-I) \\
     & $E_{cm}=3.773$ GeV & $E_{cm}=3.771$ GeV & $E_{cm}=3.774$ GeV \\
\hline
$K^-\pi^+$ & $0.27 \pm 0.02$ & $0.24\pm0.02$   & $0.25\pm0.05$ \\
$K^-\pi^+\pi^+\pi^-$ &
   $0.55 \pm 0.05$ & $0.68\pm0.11$  & $0.36\pm0.10$  \\
$K^-\pi^+\pi^+$ &
   $0.47 \pm 0.04$ & $0.38\pm0.05$  & $0.36\pm0.06$ \\
\hline \hline
\end{tabular}
\end{scriptsize}
\end{center}
\end{table}
A comparison of our measured $\sigma_D\times B$ with that measured by
MARK-II~\cite{mark2_dxsct} and  MARK-I~\cite{mark1_dxsct}
is given in Table III. 
The observed cross section for
$D^+D^-$ production is obtained by dividing the 
$\sigma_{D\bar D} \times B$
by branching fraction quoted from PDG\cite{PDG02}, which gives,
\begin{equation}
     \sigma_{D^+  {D^-} }^{\rm obs} = (2.56 \pm 0.08 \pm 0.26)~~ {\rm nb}.
\end{equation}
The observed cross sections for $D^0 {\bar D}^0$ production are
$ \sigma_{D^0 \bar {D^0} }^{\rm obs} = (3.50 \pm 0.11)~{\rm nb}$
and $\sigma_{D^0 \bar {D^0} }^{\rm obs} = (3.66 \pm 0.13)~{\rm nb}$, 
which are determined from the analysis of the singly tagged modes of
$D^0 \rightarrow K^-\pi^+$ and
$D^0 \rightarrow K^- \pi^+\pi^+\pi^-$ respectively; where the
errors are statistical.
Averaging the two observed cross sections for $D^0 {\bar D}^0$
production gives the average of the observed cross section 
for $D^0 {\bar D}^0$ production to be
\begin{equation}
   \sigma_{D^0 \bar {D^0} }^{\rm obs} = (3.58 \pm 0.09 \pm 0.31)~~ {\rm nb},
\end{equation}
\noindent
where the first error is statistical and second systematic which
is estimated based on the averaged three charged tracks in the two modes
of the single tags.
Adding the observed cross sections of the neutral and charged modes together
gives the observed cross section for $D \bar D$ production to be
\begin{equation}
   \sigma_{D \bar {D} }^{\rm obs} = (6.14 \pm 0.12 \pm 0.50)~~ {\rm nb},
\end{equation}
\noindent
where the first error is statistical and the second systematic.
In the estimation of the systematic error of the
$\sigma_{D\bar{D}}^{\rm obs}$, sources of systematic uncertainty are segregated
into components that are common or independent for $D^0$ and $D^+$
measurements. The common components are the uncertainty in the measured
integrated luminosity, the uncertainty in tracking and the uncertainty in
particle identification. Since the absolute branching fraction scale for
$D^+ \rightarrow K^-\pi^+\pi^+$ and $D^0 \rightarrow K^-\pi^+\pi^+\pi^-$
depend on the branching fraction scale for $D^0 \rightarrow K^-\pi^+$, 
the total percentage uncertainty for the two channel branching fractions
($6.6\%$ and $4.1\%$) are split into a common component 
that matches the percentage uncertainty for the 
$D^0 \rightarrow K^-\pi^+$ branching fraction ($2.3\%$) and
independent components ($6.2\%$ and $3.4\%$). All other
systematic uncertainties are treated as independent and added in quadrature.
The common uncertainties are added linearly.
\begin{table}
\caption{Comparison of the observed cross section with that measured
by MARK-III~\cite{mark3_dbltag} experiment.}
\begin{center}
\begin{tabular}{ccc}
\hline \hline
    & $\sigma^{\rm obs}_{D \bar D}$~~[nb] & 
                   $\sigma^{\rm obs}_{D \bar D}$~~ [nb] \\
    & (This experiment) & (MARK-III)  \\
    & $E_{cm}=3.773$ GeV  & $E_{cm}=3.768$ GeV \\
\hline
$\sigma_{D^0 {\bar D}^0}$ & $3.58 \pm 0.09 \pm 0.31$  &
   $2.90 \pm 0.25 \pm 0.30$  \\
$\sigma_{D^+D^-}$ & $2.56 \pm 0.08 \pm 0.26$ &  $2.10\pm0.30\pm0.15$ \\
\hline
\hline
\end{tabular}
\end{center}
\end{table}

As a comparison, Table IV lists the observed cross
sections for $D^0\bar D^0$ and $D^+D^-$ production at the c.m.
energies of 3.773 GeV and 3.768 GeV, which were measured by this experiment and
MARK-III~\cite{mark3_dbltag}.

\section{Radiative Corrections}
In any $e^+e^-$ colliding beam experiment, the electron (positron) always
radiates at the interaction point because of the potential of the
positron (electron). 
Since this radiation (Bremsstrahlung) carries
energy away, 
the actual center-of-mass energy for the $e^+e^-$ annihilation is reduced by
Bremsstrahlung to $\sqrt{s(1-x)}$, where $xE_{\rm beam}$ is the total energy
of the emitted photons.
The Bremsstrahlung is principally responsible for the distortions
to the tree level resonance line shape, while the self energy
of the electron and positron and the vertex corrections to the initial state
affect the overall factors to change the scale of the cross section. 
All of these corrections are called Initial State Radiation (ISR) corrections.
The tree level cross section for $D \bar D$ production at the energy
of 3.773 GeV can be obtained by correcting the
observed cross section for the effects of the ISR and
vacuum polarization. 

The observed cross section, $\sigma^{\rm obs}$,
at the nominal energy $\sqrt{s}$ can be written as a convolution of the
Born cross section $\sigma^B(s(1-x))$ and
a sampling function $f(x,s)$,
\begin{equation}
  \sigma^{\rm obs}(s)= \int^1_0 dx 
          \cdot f(x,s)\sigma^{B}(s(1-x))(1+\delta_{VP}(s(1-x))) .
\end{equation}
\noindent
The vacuum polarization correction $(1+\delta_{VP})$ includes both
leptonic and hadronic terms. It varies from charm threshold to
4.14 GeV by
less than $\pm 2\%$~\cite{bes4030}. In this data analysis, we treat it as a constant of
\begin{equation}
    (1+\delta_{VP})=1.047\pm0.024.
\end{equation}
Since we are interested in the $\psi(3770)$ resonance
in this analysis, we take the $\sigma^B$ to be the bare
Breit-Wigner cross section
\begin{equation}
\sigma^B(E) = \frac{12 \pi \Gamma^0_{ee}\Gamma_{\rm tot}(E)}
{{(E^2-M^2)^2 + M^2\Gamma^2_{\rm tot}(E)}},
\end{equation}
where $\Gamma^0_{ee} = \Gamma_{ee}/(1+\delta_{\rm {vp}})$,
$M$ and $\Gamma_{ee}$ are the mass and leptonic width
of the $\psi(3770)$ resonance respectively;
$E$ is the center-of-mass energy;
$\Gamma_{\rm tot}(E)$ is chosen to be energy dependent and normalized to
the total width $\Gamma_{\rm tot}$
at the peak of the resonance~\cite{PDG02}\cite{mark2}.
The $\Gamma_{\rm tot}(E)$ is defined as
\begin{equation}
 \Gamma_{\rm tot}(E) =  \Gamma_0  \frac{  \frac{p^3_{D^0}}{1+(r {p_{D^0}})^2}
                          + \frac{p^3_{D^{\pm}}}{1+(r {p_{D^{\pm}}})^2} }
                 {  \frac{{p^0}^3_{D^0}}{1+(r {{p^0}_{D^0}})^2}
                          + \frac{{p^0}^3_{D^{\pm}}}{1+(r
{{p^0}_{D^{\pm}}})^2} },
\end{equation}
where $p^0_D$ is the momentum of the $D$ mesons produced at the
peak of $\psi(3770)$, $p^{ }_D$ is the momentum of the $D$ mesons
produced at the c.m. energy $\sqrt{s}$, $\Gamma_0$ is the width of
the $\psi(3770)$ at the peak, 
and $r$ is the interaction radius which was set to be 0.5 fm.
In the calculation of the Born order cross section, the $\psi(3770)$
resonance parameters $M=3769.9\pm 2.5$ MeV; $\Gamma_0=23.6\pm2.7$ MeV
and $\Gamma_{ee}=0.26 \pm 0.04$ keV~\cite{PDG02}
were used.

In the structure function  approach  introduced  by
Kuraev and 
Fadin~\cite{kuraev}\cite{altarelli}, 
the sampling
function can be written as
\begin{center}
\begin{equation}
f(x,s)\;=\;\beta x^{\beta-1}\delta^{V+S}+\delta^{H} ,
\end{equation}
\end{center}
where $\beta$ is the electron equivalent radiator thickness,
\begin{equation}
\beta\;=\;\frac{2\alpha}{\pi} \left(\ln \frac{s}{m^{2}_{e}}-1\right) ,
\end {equation}
\begin{equation}
\delta^{V+S}\;=\;1+\frac{3}{4}\beta+\frac{\alpha}{\pi}
\left(\frac{\pi^{2}}{3}-\frac{1}{2}\right)+
\frac{\beta^2}{24}\left(\frac{1}{3}\ln \frac{s}{m^2_e}+2\pi^2-\frac{37}{4}\right),
\end{equation}
\begin{equation}
\delta^{H}\;=\;\delta^{H}_{1}+\;\delta^{H}_{2} ,
\end{equation}
\begin{equation}
\delta^{H}_{1}\;=\;-\beta\left(1-\frac{x}{2}\right) ,
\end{equation}

\begin{equation}
\delta^{H}_{2}\;=\;\frac{1}{8}\beta^{2}\left[4(2-x)\ln\frac{1}{x}-
\frac{1+3(1-x)^{2}}{x}\ln(1-x)-6+x\right] .
\end{equation}

\noindent
In the above formula, $m_e$ is the electron mass and $\alpha$
is the fine structure constant.
    The $\psi(3770)$ width ($\sim 24$ MeV) is much large than the energy
spread ($\sim 1.37 $ MeV) of BEPC. So
the effect of the beam energy spread
on the cross section could be ignored. The $\psi(3770)$
is generally assumed to decay exclusively into $D \bar D$.
Taking these considerations, 
the observed cross section of Equation (7)
should be replaced by
\begin{equation}
  \sigma^{\rm obs}(s)=(1+\delta_{VP}) \int^{1-4M_D^2/s}_0  dx 
              \cdot f(x,s)\sigma^{B}(s(1-x))
\end{equation}
\noindent
in calculation of the radiative corrections.
The correction factor for the radiative effects is given by
\begin{equation}
 g = \frac{\sigma^{\rm obs}}{\sigma^{B}}. 
\end{equation}
\noindent
Figure 2 shows the factor of the radiative corrections as a function
of the nominal center-of-mass energy. At the center-of-mass energy 
$\sqrt{s}=3.773~ \rm GeV$, the factor is
\begin{equation}
 g = 0.779\pm 0.031, 
\end{equation}
\noindent
where the error is the uncertainty arising from the errors on the $\psi(3770)$
resonance parameters. The uncertainty is mainly due to the error of the
mass of the resonance.
\begin{figure}
\includegraphics[width=9.0cm,height=9.0cm]
{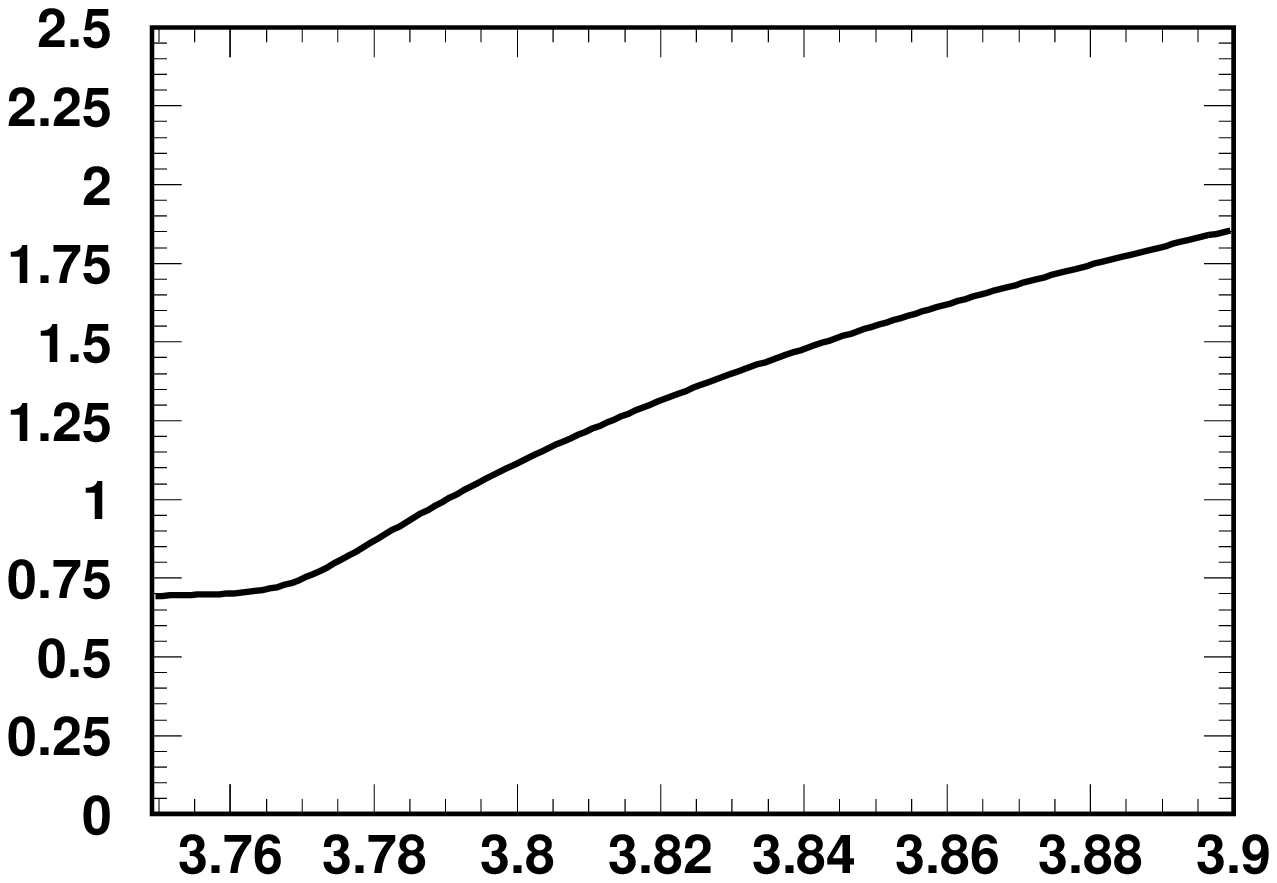}
\put(-170.0,-4.0){\bf\large {Nominal $E_{cm}$~~~~[GeV]}}
\put(-260.0,170.0){\Large{$g$}}
\caption{The factor of radiative corrections as a function
of the nominal center-of-mass energy.}
\end{figure}

\section{Cross section for $D \bar D$ production}

    The tree level cross section for $D \bar D$ production is
obtained by dividing the observed cross section by the factor
of the radiative corrections. At $\sqrt{s}=3.773$ GeV, 
the charged, neutral and total tree level $D$ pair production cross sections are
\begin{equation}
    \sigma_{D^0 {\bar D}^0} = (4.60 \pm 0.12 \pm 0.45)~~\rm nb, 
\end{equation}
\begin{equation}
    \sigma_{D^+D^-} = (3.29 \pm 0.10 \pm 0.37)~~\rm nb, 
\end{equation}
\noindent
and
\begin{equation}
    \sigma_{D \bar D} = (7.88 \pm 0.15 \pm 0.74)~~\rm nb, 
\end{equation}
\noindent
where the first error is statistical and the second systematic which
include the uncertainty in the factor of the radiative corrections.
These results are compared to the coupled-channel model prediction
in Table V.
\begin{table}
\caption{Comparison of tree level cross section measurements with
prediction of the coupled-channel model at $\sqrt{s}=3.773$ GeV.}
\begin{center}
\begin{tabular}{ccc}
\hline \hline
    & Experiment & coupled-channel Model  \\
\hline
$\sigma_{D^0 {\bar D}^0}$ & $4.60 \pm 0.12 \pm 0.45$ nb  &
   $1.80$ nb \\
$\sigma_{D^+D^-}$ & $3.29 \pm 0.10 \pm 0.37$ nb &  $1.28$ nb \\
$\sigma_{D\bar D}$ & $7.88 \pm 0.15 \pm 0.74$ nb & $3.08$ nb \\
\hline
\hline
\end{tabular}
\end{center}
\end{table}

\section{Measurement of $R_D$ and $R$}

The tree level cross section for $\mu^+\mu^-$ production in QED is
given by
\begin{equation}
          \sigma_{e^+e^- \rightarrow \mu^+\mu^-} = 
           \frac{86.8~ {\rm nb}}{E^2_{cm}},
\end{equation}
\noindent
where the $E_{cm}$ is the center-of-mass energy in GeV. A
measurement of $R_D$~\cite{coles} is obtained by dividing
$2\sigma_{D \bar D}$ by the tree level muon pair cross
section, which gives
\begin{equation}
      R_D = 2.58 \pm 0.05 \pm 0.24.
\end{equation}
\noindent
BES-II experiment measured $R_{uds}$\cite{bes2R}, which is the ratio of the tree level
light hadron (containing the u,d and s light quarks) cross section over that
for $\mu^+\mu^-$ production in the energy region from 2.0 to 3.0 GeV.
Theoretical expectation is that $R_{uds}$ is approximately independent
of center-of-mass energy in this region~\cite{bes4030}. 
Fitting to the $R_{uds}$ values at
9 energy points in the energy region, we obtain
$R_{uds}=2.26\pm0.14$. Assuming that $\psi(3770)$ decays exclusively into 
$D \bar D$, the value of $R$ is evaluated using $R=R_D/2 + R_{uds}$,
which gives
\begin{equation}
R=3.55 \pm 0.03 \pm 0.18. 
\end{equation}
\section{Summary}

    In summary, using the 17.3 $\rm pb^{-1}$ of data collected 
with the BES-II detector at BEPC at center-of-mass energy 
$\sqrt{s}=3.773$ GeV, the observed cross sections for 
$D^0 {\bar D}^0$, $D^+D^-$ and $D\bar D$ production have
been measured. Those are 
$\sigma_{D^0 \bar {D^0} }^{\rm obs} = (3.58 \pm 0.09 \pm 0.31)~{\rm nb}$,
$\sigma_{D^+  {D^-} }^{\rm obs} = (2.56 \pm 0.08 \pm0.26)~{\rm nb}$ and
$\sigma_{D \bar {D} }^{\rm obs} = (6.14 \pm 0.12 \pm 0.50)~{\rm nb}$.
The tree level cross sections for the 
$D^0 {\bar D}^0$, $D^+D^-$ and $D\bar D$ production 
are determined to be
$\sigma_{D^0 {\bar D}^0} = (4.60 \pm 0.12 \pm 0.45)~\rm nb$,
$\sigma_{D^+D^-} = (3.29 \pm 0.10 \pm 0.37)~\rm nb$ and
$\sigma_{D \bar D} = (7.88 \pm 0.15 \pm 0.74)~\rm nb$,
which are about a factor 2.5 times larger than that predicted
by the coupled-channel model. 
Using the measured $R_{uds}$
in the energy region from 2.0 to 3.0 GeV from BES-II experiment
and assuming that $\psi(3770)$ only decays to $D \bar D$,
the total tree level cross section 
for inclusive hadronic event production at 3.773
GeV is obtained to be $R=3.55 \pm 0.03 \pm 0.18$.

\vspace{3mm}

\begin{center}
{\small {\bf ACKNOWLEDGEMENTS}}
\end{center}

\vspace{-0.4cm}

   The BES collaboration thanks the staff of BEPC for their hard efforts.
This work is supported in part by the National Natural Science Foundation
of China under contracts
Nos. 19991480,10225524,10225525, the Chinese Academy
of Sciences under contract No. KJ 95T-03, the 100 Talents Program of CAS
under Contract Nos. U-11, U-24, U-25, and the Knowledge Innovation Project
of CAS under Contract Nos. U-602, U-34(IHEP); by the
National Natural Science
Foundation of China under Contract No.10175060(USTC),and
No.10225522(Tsinghua University).

\end{document}